\documentclass[
draft            
  ]
  {aipproc}

\layoutstyle{6x9}

\usepackage{epsfig}
\def\etal{{\it et al.}}

\begin{document}
\begin{flushright}SUHEP-05-2006 \end{flushright}
\begin{flushright}April, 2006~~~~~~~~~~ \end{flushright}
\title{Experimental Status of $B$ Physics}



\author{Sheldon Stone}{
  address={Physics Department, Syracuse University, Syracuse, N. Y., USA
  13244-1130}
  }

\begin{abstract}
A short summary is given of the current status of $B$ physics.
Reasons for physics beyond the Standard Model are discussed.
Constraints on New Physics are given using measurements of $B$
mixing, $B_S$ mixing, and CP violation, along with $|V_{ub}|$.
Future goals, and upcoming new experiments are also mentioned.
\end{abstract}

\maketitle


\section{Introduction}
``New Physics" (NP) refers to physics beyond the ``Standard
Model," the paradigm that we have constructed to explain our
current high energy physics data \cite{SM}. We know, however, that
NP is required to explain certain global phenomena including the
Baryon asymmetry of the Universe, without which we could not
exist, or the ``Dark Matter," found first by Zwicky studying
rotation curves of galaxies \cite{Zwicky}. An even more mysterious
phenomena called ``Dark Energy" may also have a connection to
particle physics experiments \cite{Trodden}, perhaps via ``Extra
Dimensions" \cite{ExtraD}.
 The fundamental goals of $B$ decay studies are to discover, or
help interpret, NP found elsewhere. Additional goals include
measuring  ``fundamental constants" revealed to us by studying
Weak interactions and understand the theory of strong
interactions, QCD, necessary to interpret our measurements.

\subsection{Baryogenesis}

When the Universe began with the Big Bang, there was an equal amount
of matter and antimatter. Now we have mostly matter. How did it
happen? A. Sakharov gave three necessary conditions: Baryon (B)
number violation, departure from thermal equilibrium, and C and CP
violation \cite{Sakh}. (The operation of Charge Conjugation (C)
takes particle to anti-particle and Parity (P) takes a vector
$\overrightarrow{r}$ to $-\overrightarrow{r}$.)

These criteria are all satisfied by the Standard Model. B is
violated in Electroweak theory at high temperature, though baryon
minus lepton number is conserved; in addition we need quantum
tunneling, which is powerfully suppressed at the low temperatures
that we now have. Non-thermal equilibrium is provided by the
electroweak phase transition. C and CP are violated by weak
interactions. However the violation is too small. The ratio of the
number of baryons to the number of photons in the Universe needs
to be $\sim 6\times 10^{-10}$, while the SM can provide only $\sim
10^{-20}$. Therefore, there must be new physics.

\subsection{The Hierarchy Problem}
Definition from the WIKIPEDIA encyclopedia \cite{wiki}: ``In
theoretical physics, a hierarchy problem occurs when the
fundamental parameters (couplings or masses) of some Lagrangian
are vastly different (usually larger) than the parameters measured
by experiment. This can happen because measured parameters are
related to the fundamental parameters by a prescription known as
renormalization. Typically the renormalized parameters are closely
related to the fundamental parameters, but in some cases, it
appears that there has been a delicate cancellation between the
fundamental quantity and the quantum corrections to it."

Our worry is why the Planck scale at $\sim 10^{19}$ GeV is so much
higher than the scale at which we expect to find the Higgs Boson,
$\sim$100 GeV. We expect the explanation lies in physics beyond the
Standard Model.

\section{The Basics: Quark Mixing and the CKM Matrix}
\label{CKM} The CKM matrix parameterizes the mixing between the
mass eigenstates and weak eigenstates as couplings between the
charge +2/3 and -1/3 quarks. I use here the Wolfenstein
approximation \cite{Wolf} good to order $\lambda^3$ in the real
part and $\lambda^4$ in the imaginary part:
\begin{equation}
V_{CKM}=  \label{eq:CKM} \left(\begin{array}{ccc}
1-\lambda^2/2 &  \lambda & A\lambda^3(\rho-i\eta(1-\lambda^2/2)) \\
-\lambda &  1-\lambda^2/2-i\eta A^2\lambda^4 & A\lambda^2(1+i\eta\lambda^2) \\
A\lambda^3(1-\rho-i\eta) &  -A\lambda^2& 1
\end{array}\right).
\end{equation}

In the Standard Model $A$, $\lambda$, $\rho$ and $\eta$ are
fundamental constants of nature like $G$, or $\alpha_{EM}$; $\eta$
multiplies $i$ and is responsible for all Standard Model CP
violation. We know $\lambda$=0.226, $A \sim$0.8 and we have
constraints on $\rho$ and $\eta$.

\begin{figure}
\centerline{\epsfig{figure= 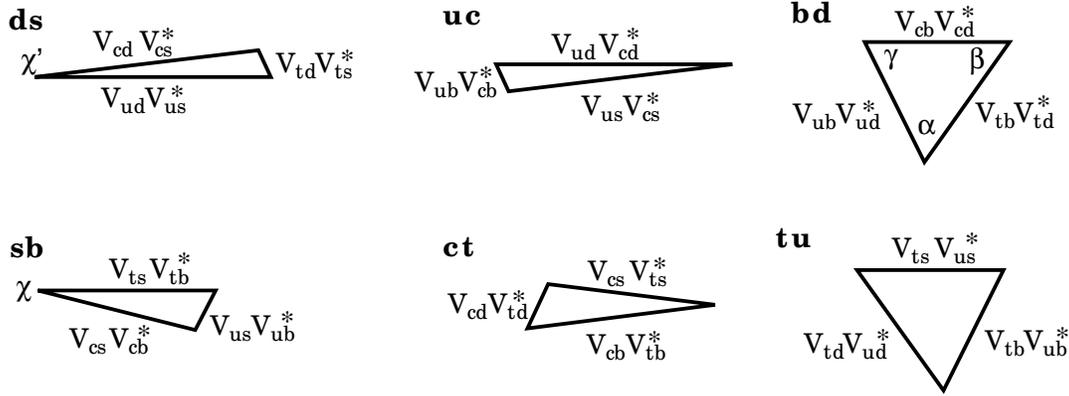,height=2.1in}}
\caption{The 6 CKM triangles resulting from applying unitarity
constraints to the indicated row and column. The CP violating
angles are also shown.}
\label{six_tri}       
\end{figure}

Applying unitarity constraints allows us to construct the six
independent triangles shown in Fig.~\ref{six_tri}. Another basis
for the CKM matrix are four angles labelled as $\chi$ (sometimes
called $\beta_S$), $\chi'$ and any two of $\alpha$, $\beta$ and
$\gamma$ since $\alpha +\beta +\gamma =\pi$ \cite{akl}. (These
angles are also shown in Fig.~\ref{six_tri}.)

\begin{figure}
\centerline{\epsfig{figure=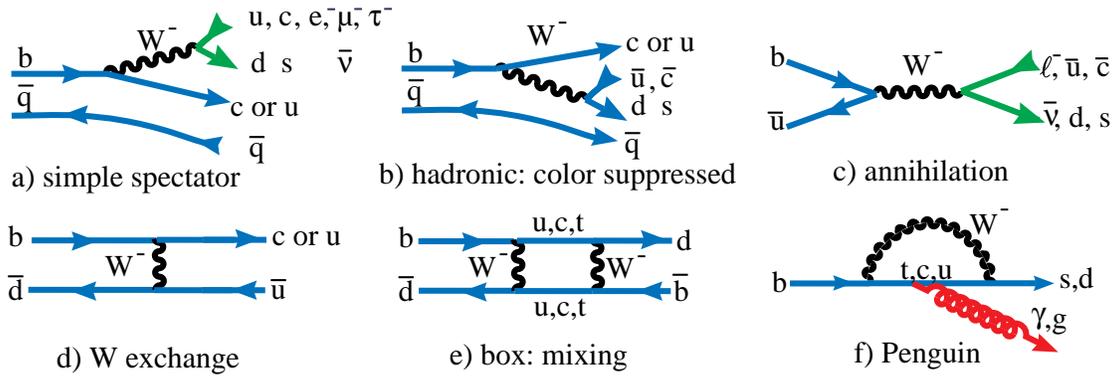,height=2.0in}}
\caption{Some $B$ decay diagrams.}
\label{Bdiagrams2}       
\end{figure}

$B$ meson decays can occur through various processes. Some decay
diagrams are shown in Fig.~\ref{Bdiagrams2}. The simple spectator
diagram is dominant. Semileptonic decays, discussed next, proceed
through this diagram.

\section{Semileptonic Decays and Lifetimes}

These are the simplest decays to describe theoretically. The
transformation of the virtual $W^-$ to a lepton-antineutrino pair
proceeds through the axial-vector current just as in pion decay.
Because of their relative simplicity, semileptonic decays are used
to probe the $b\to c$ and $b\to u$ transitions. The overall
semileptonic branching ratio, ${\cal{B}}_{SL}$ is defined as
${\cal{B}}(\to X e^-\bar{\nu})$ equal to ${\cal{B}}(\to X
\mu^-\bar{\nu})$ and has a measured value of (10.2$\pm$0.9)\% and
(10.5$\pm$0.8)\%, for $B^-$ and $\overline{B}^0$ mesons,
respectively. The average for these mesons is much better measured
as (10.87$\pm$0.17)\% \cite{PDG}.

The rather long average $B$ lifetime, $\sim$1.5 ps is an important
aspect of $B$ decays and is a crucial property allowing for more
precise measurements of CP violation and other properties. The
lifetime ratio $\tau_{B^-}/\tau_{\overline{B}^0}$=
1.071$\pm$0.0009 clearly demonstrates a longer, but not much
longer lifetime for charged versus neutral $B$ mesons.

Measurements of the CKM matrix elements $|V_{cb}|$ and $|V_{ub}|$
have been made using both exclusive decays to specific final
states, such as $B\to D^*\ell^-\bar{\nu}$ and inclusive final
states. Values have been compiled by the Heavy Flavor Averaging
Group \cite{HFAG}. $|V_{cb}|$ is measured to be 0.038$\pm$0.001
from exclusive decays using Heavy Quark Effective Theory (HQET)
\cite{PDG}. Inclusive decays have also been used and good
precision has been achieved, although the accuracy depends
critically on whether or not the assumption of ``duality" is
indeed correct. Measurements of $|V_{ub}|$ have also been made
also using exclusive and inclusive decays. It is in the range of
$3-4\times 10^{-3}$. The main uncertainties are theoretical since
there isn't a firm theoretical basis similar to HQET that can be
used. The combination $|V_{ub}/V_{cb}|\approx
\lambda^2\sqrt{\rho^2+\eta^2}$.

\section{Current $B$ Decay Experiments}

\subsection{Current {$e^+e^-$} Experiments}

The BaBar and Belle collaborations both work at $e^+e^-$
colliders, with asymmetric energies, at a center-of-mass energy
equal to the mass of the $\Upsilon(4S)$ resonance. Here there is
almost equal production of both $B^-B^+$ and $\overline{B}^0B^0$
pairs, totaling 1 nb of cross section on top of 3 nb of background
quark-antiquark production. The asymmetric energies are necessary
to boost the $B^0$ mesons so that CP violation measurements can be
made; the time integrated asymmetries would otherwise vanish as
they are in $J^{PC}=1^{--}$ states \cite{Bigi-Sanda}. The boost,
however, is small so the decay time resolution is only $\sim$900
fs r.m.s.

CLEO and ARGUS collected data on the $\Upsilon$ resonances using
symmetric $e^+e^-$ machines. CLEO is now concentrating in studying
charm meson decays at lower energies. It is also worth noting that
many $e^+e^-$ experiments have provided a wealth of interesting
data including the aforementioned ones and experiments at LEP
(operating at the $Z^0$ resonance) and the PEP and PETRA machines
(operating near 30 GeV).

Both CLEO and Belle have taken data at the $\Upsilon(5S)$ resonance.
CLEO has determined the $B_S$ fraction $\sim$16\% of the 0.3 nb
$b\overline{b}$ cross-section, about 1/20 the production rate at the
$\Upsilon(4S)$\cite{CLEO5S}. Not only is the yield small but the
proper time resolution is not sufficient to allow time dependent CP
violation measurements.

\subsection{Current Hadron Collider Experiments}
The CDF and D0 experiments at the Fermilab Tevatron are designed
to study high energy phenomena, such as finding the top-quark and
Higgs boson. However, they have some $b$ capabilities and are well
suited to study the $B_S$ meson, which cannot be studied with
$e^+e^-$ colliders. The most important measurement that may be
within reach of these experiments is that of $B_S$ mixing.
Production of $b$-flavored hadrons is a large 100 $\mu$b at the 2
TeV center-of-mass energy of the Tevatron. Unfortunately the
detectors are as not fully equipped as dedicated heavy flavor
experiments. They lack the excellent particle identification and
crystal based electromagnetic calorimetry of the state-of-the-art
$e^+e^-$ experiments. They do, however, have good $\sim$100 fs
decay time resolution \cite{Giurgiu}.

\section{{$B_d$} and {$B_S$} Mixing}

A diagram for $B_d$ mixing is shown in Fig.~\ref{Bdiagrams2}(e). For
$B_S$ mixing just replace the $d$ quarks with $s$ quarks. The flavor
eigenstates, degenerate in pure QCD mix under the weak interactions.
Designating the base states as $\{|1>,|2>\}\equiv
\{|B^0>,|\overline{B}^0>\}$, the Hamiltonian is
\begin{equation}
H=M=-{i\over 2}\Gamma = \left(\begin{array}{cc}
M_{11}&M_{12} \\
M^*_{12} &  M_{22}
\end{array}\right)
=-{i\over 2} \left(\begin{array}{cc}
\Gamma_{11}&\Gamma_{12} \\
\Gamma^*_{12} &  \Gamma_{22}
\end{array}\right)~.
\end{equation}

\begin{figure}
\centerline{\epsfig{figure= 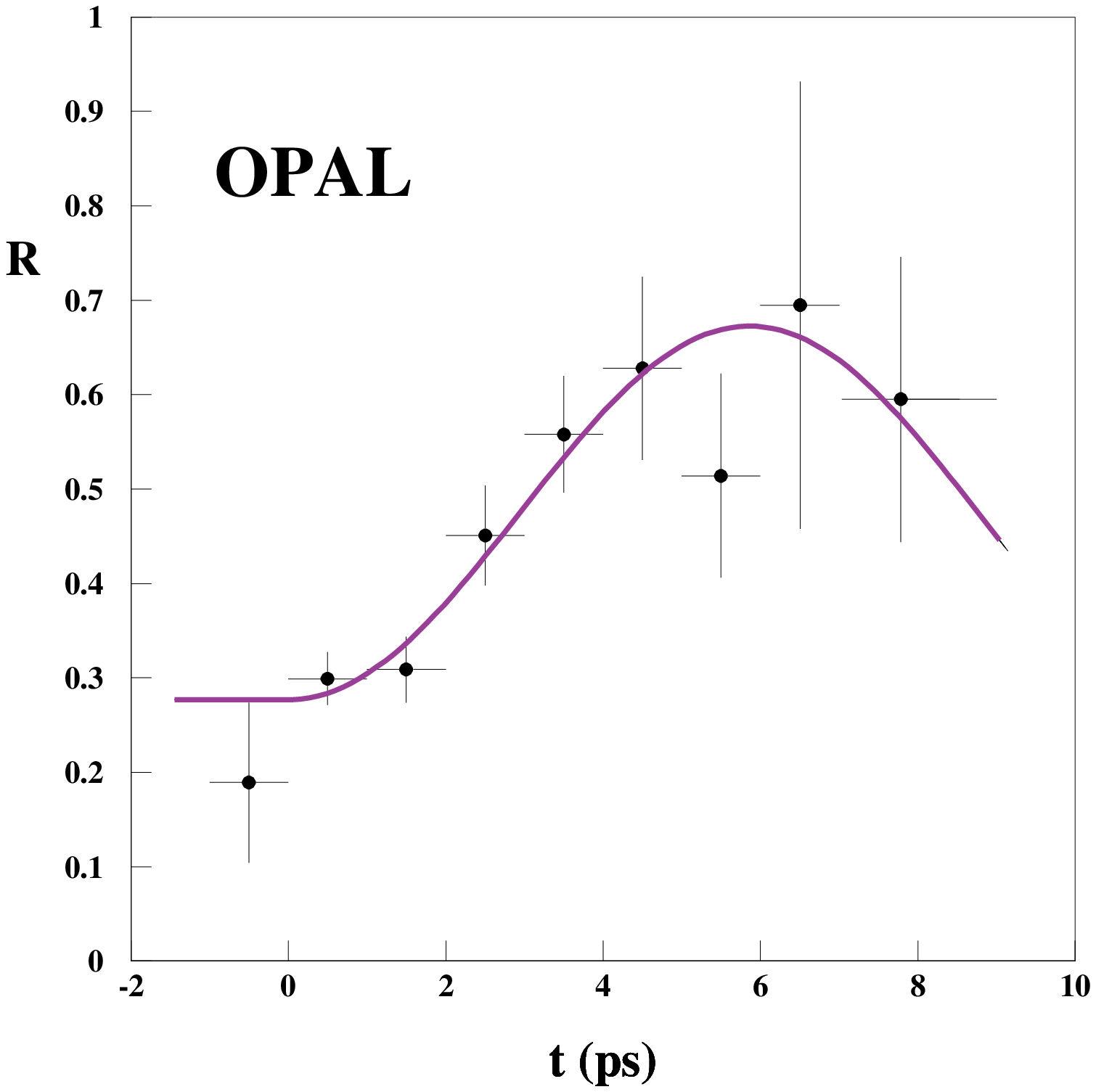,height=2.2in}}
\caption{The probability
$R=\left(B^0\to\overline{B}^0\right)/\left(B^0\to {B}^0\right)$ as
a function of time from the OPAL experiment.}
\label{opal_mix_d}       
\end{figure}

Diagonalizing the matrix we find the mass difference $\Delta m =
m_{B_H}-m_{B_L}=2|M_{12}| .$ For $B_d$ we predict $\Delta\Gamma$
$\sim$0.\footnote{This is because the fraction of final states
that of the same CP parity that both $B^0$ and $\overline{B}^0$
can decay into is very small. This is not the case for $B_S$.} The
probability for a $B^0$ meson to appear as a $\overline{B}^0$ as a
function of time is given by $0.5\Gamma e^{-\Gamma
t}\left[1+\cos(\Delta m t)\right]$. $R$ is often defined as the
ratio $\left(B^0\to\overline{B}^0\right)/\left(B^0\to
{B}^0\right)$. $B_d$ mixing was first discovered by the ARGUS
experiment \cite{ARGUS-mix}. (There was a previous measurement by
UA1 indicating mixing for a mixture of $B_d^0$ and $B_s^0$
\cite{Albajar-1987}.) At the time it was quite a surprise, since
$m_t$ was thought to be in the 30 GeV range. It is usual to define
$R$ as probability for a $B^0$ to materialize as a
$\overline{B}^0$ divided by the probability it decays as a $B^0$.
An early mixing result from OPAL is shown in Fig.~\ref{opal_mix_d}
\cite{Akers-1995}.
The world average value for $\Delta m_d$ is a very precise
0.509$\pm$0.005 ps$^{-1}$ \cite{PDG}. The measurement is dominated
by the BaBar and Belle experiments.

The probability of mixing is given by \cite{Gaillard 1974} as
\begin{equation}
x\equiv \frac{\Delta m}{\Gamma}={G_F^2\over
6\pi^2}B_Bf_B^2m_B\tau_B|V^*_{tb}V_{td}|^2m_t^2 F{\left(m^2_t\over
M^2_W\right)}\eta_{QCD}, \label{eq:Bdmix}
\end{equation}
where $B_B$ is a parameter related to the probability of the $d$ and
$\bar{b}$ quarks forming a hadron and must be estimated
theoretically, $F$ is a known function which increases approximately
as $m^2_t$, and $\eta_{QCD}$ is a QCD correction, with value about
0.8. By far the largest uncertainty arises from the unknown decay
constant, $f_B$. In principle $f_B$ can be measured. The decay rate
of the annihilation process $B^-\to\ell^-\overline{\nu}$ is
proportional to the product of $f_B^2|V_{ub}|^2$. This is a very
difficult process to measure, and even if this were done, the
uncertainty on $V_{ub}$ will lead to an imprecise result. Our
current best hope is to rely on unquenched lattice QCD which can use
the measurements of the analogous $D^+\to\mu^+\nu$ decay as check.
These checks are currently in progress at CLEO-c \cite{Artuso}.
Since
\begin{equation}
 |V^*_{tb}V_{td}|^2\propto |(1-\rho-i\eta)|^2=(\rho-1)^2+\eta^2,
\label{eq:mixrhoeta}
\end{equation}
measuring mixing gives a circle centered at (1,0) in the $\rho -
\eta$ plane. This could in principle be a very powerful
constraint. Unfortunately, the parameter $B_B$ is not
experimentally accessible and $f_B$ must be calculated; the errors
on the calculations are quite large.

$B^0_s$ mesons can mix in a similar fashion to $B^0_d$ mesons. The
diagram in Fig.~\ref{Bdiagrams2}(e) is modified by substituting
$s$ quarks for $d$ quarks, thereby changing the relevant CKM
matrix element from $V_{td}$ to $V_{ts}$. Measuring $x_s$ allows
us to use ratio of $x_d/x_s$ to provide constraints on the CKM
parameters $\rho$ and $\eta$. We still obtain a circle in the
($\rho ,\eta$) plane centered at (1,0):
\begin{eqnarray}
\left|V_{td}\right|^2&=&A^2\lambda^4\left[(1-\rho)^2+\eta^2\right] \\
{\left|V_{td}\right|^2 \over
\left|V_{ts}\right|^2}&=&(1-\rho)^2+\eta^2 \nonumber ~~.
\end{eqnarray}
Now however we must calculate only the SU(3) broken ratios
$B_{B_d}/B_{B_s}$ and $f_{B_d}/f_{B_s}$.

$B^0_s$ mixing has been searched for at LEP, the Tevatron, and the
SLC. A combined analysis has been performed. The probability,
${\cal{P}}(t)$ for a $B_s$ to oscillate into a $\overline{B}_s$ is
given as
\begin{equation}
{\cal{P}}(t)\left(B_s\to\overline{B}_s\right)={1\over 2}\Gamma_s
e^{-\Gamma_s t} \left[1+\cos\left(\Delta m_s t\right)\right]~~,
\end{equation}
where $t$ is the proper time.

\begin{figure}[htb]
\centerline{\epsfig{figure=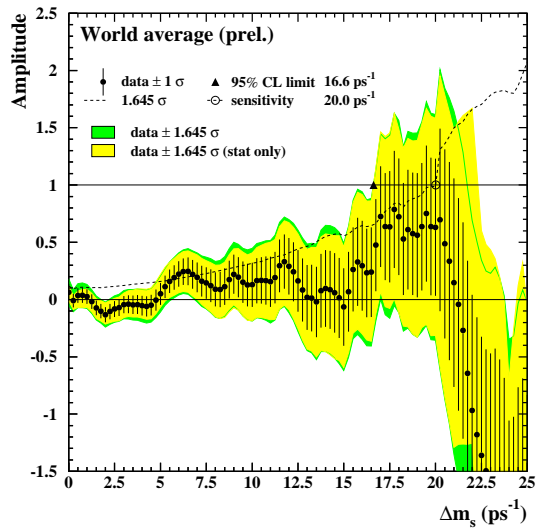,width=3in}}
\caption{\label{Bs_mix_sum} Combined experimental values of the
amplitude $A$ versus the test  frequency $\omega = \Delta m_s$ as
defined in equation~\ref{eq:Bs}. The inner (outer) envelopes give
the 95\% confidence levels using statistical (statistical and
systematic) errors. The ``sensitivity" shown at 20.0 ps$^{-1}$ is
the likely place a 95\% c.l. upper limit could be set.}
\end{figure}

To combine different experiments a framework has been established
where each experiment finds a amplitude $A$ for each test
frequency $\omega$, defined as
\begin{equation}
{\cal{P}}(t)={1\over 2}\Gamma_s e^{-\Gamma_s t}
\left[1+A\cos\left(\omega t\right)\right]~~. \label{eq:Bs}
\end{equation}
Fig.~\ref{Bs_mix_sum} shows the world average measured amplitude $A$
as a function of the test frequency $\omega=\Delta m_s$ \cite{HFAG}.
For each frequency the expected result is either zero for no mixing
or one for mixing. No other value is physical, although measurement
errors admit other values. The data do indeed cross one at a $\Delta
m_s$ of 16 ps$^{-1}$, however here the error on $A$ is about 0.6,
precluding a statistically significant discovery. The quoted upper
limit at 95\% confidence level is 16.6 ps$^{-1}$. This is the point
where the value of $A$ plus 1.645 times the error on $A$ reach one.
Also, one should be aware that all the points are strongly
correlated.

\begin{figure}[htb]
\includegraphics[width=0.50\textwidth]{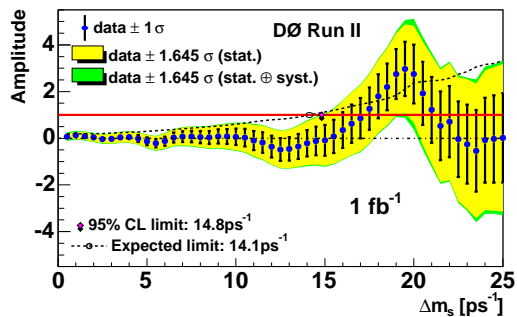}
\caption{\label{prl_fig3} $B^0_S$ oscillation amplitude as a
function of oscillation frequency, $\Delta m_s$ from D0.  The red
(solid) line shows the $\mathcal{A}=1$ axis for reference.  The
dashed line shows the expected limit including both statistical and
systematic uncertainties.  }
\end{figure}

As this work was being completed the D0 experiment announced that
they had limited the $\Delta m_s$ between 17 ps$^{-1}$ and 21
ps$^{-1}$ at 90\% confidence level \cite{D0-BS}.
Fig.~\ref{prl_fig3} shows their amplitude analysis results.
Clearly the significance of the result, although limited, relies
on seeing an amplitude in excess of the expected value of 1, in
fact, nearly at 3.  Further data will be needed to confirm this
result. The inferred values of $\rho$ and $\eta$ are within the
range expected by fits to other parameters (see
Fig.~\ref{rhoeta}), and are consistent with Standard Model
expectations.

\section{CP Violation Measurements}

\subsection{Introduction}
\label{section:CPV}

CP violation can occur because of the imaginary term in the CKM
matrix, proportional to $\eta$ in the Wolfenstein representation
\cite{Bigi-Sanda}.

Decays of neutral $K$ mesons were the first to show CP violating
effects. In this decade the BaBar and Belle experiments provided
precision measurement of one of the four CP violating angles
($\beta$) and gave first measurements of two other angles
($\alpha$ and $\gamma$).

 Consider the case of a process $B\to f$
that goes via two amplitudes \cal{A} and \cal{B} each of which has
a strong part e. g. $s_{\cal{A}}$ and a weak part $w_{\cal{A}}$.
Then we have
\begin{eqnarray}
\Gamma(B\to
f)&=&\left(\left|{\cal{A}}\right|e^{i(s_{\cal{A}}+w_{\cal{A}})}
+\left|{\cal{B}}\right|e^{i(s_{\cal{B}}+w_{\cal{B}})}\right)^2 \\
\Gamma(\overline{B}\to \overline{f})&=&
\left(\left|{\cal{A}}\right|e^{i(s_{\cal{A}}-w_{\cal{A}})}
+\left|{\cal{B}}\right|e^{i(s_{\cal{B}}-w_{\cal{B}})}\right)^2 \\
\Gamma(B\to f)-\Gamma(\overline{B}\to
\overline{f})&=&2\left|{\cal{AB}}\right|
\sin(s_{\cal{A}}-s_{\cal{B}})\sin(w_{\cal{A}}-w_{\cal{B}})~~.
\end{eqnarray}

Any two amplitudes will do, though its better that they be of
approximately equal size. Thus charged $B$ decays can exhibit CP
violation as well as neutral $B$ decays. In some cases, we will
see that it is possible to guarantee that
$\left|\sin(s_{\cal{A}}-s_{\cal{B}})\right|$ is unity, so we can
get information on the weak phases. In the case of neutral $B$
decays, mixing can be the second amplitude.

\subsection{Formalism of CP Violation in Neutral $B$ Decays}

For neutral mesons we can construct the CP eigenstates
\begin{equation}
\big|B^0_1\big>={1\over \sqrt{2}}\left(\big|B^0\big>-
\big|\overline{B}^0\big>\right), ~~~~\big|B^0_2\big>={1\over
\sqrt{2}}\left(\big|B^0\big>+\big|\overline{B}^0\big>\right),~{\rm
where}
\end{equation}
\begin{equation}
CP\big|B^0_1\big>=\big|B^0_1\big>,
~~~~CP\big|B^0_2\big>=-\big|B^0_2\big>~~.
\end{equation}
Since $B^0$ and $\overline{B}^0$ can mix, the mass eigenstates are
a superposition of $a\big|B^0\big> + b\big|\overline{B}^0\big>$
which obey the Schrodinger equation
\begin{equation}
i{d\over dt}\left(\begin{array}{c}a\\b\end{array}\right)= {\cal
H}\left(\begin{array}{c}a\\b\end{array}\right)= \left(M-{i\over
2}\Gamma\right)\left(\begin{array}{c}a\\b\end{array}\right).
\label{eq:schrod}
\end{equation}
If CP is not conserved then the eigenvectors, the mass eigenstates
$\big|B_L\big>  $ and  $\big|B_H\big>$, are not the CP eigenstates
but are
\begin{equation}
\big|B_L\big> =
p\big|B^0\big>+q\big|\overline{B}^0\big>,~~\big|B_H\big> =
p\big|B^0\big>-q\big|\overline{B}^0\big>,~{\rm where}
\end{equation}
\begin{equation}
p={1\over \sqrt{2}}{{1+\epsilon_B}\over
{\sqrt{1+|\epsilon_B|^2}}},~~ q={1\over
\sqrt{2}}{{1-\epsilon_B}\over {\sqrt{1+|\epsilon_B|^2}}}.
\end{equation}
CP is violated if $\epsilon_B\neq 0$, which occurs if $|q/p|\neq
1$.

\subsubsection{CP violation for $B$ via interference of mixing and decays}

Here we choose a final state $f$ which is accessible to both $B^0$
and $\overline{B}^0$  decays. The second amplitude necessary for
interference is provided by mixing. It is necessary only that $f$
be accessible directly from either state; however if $f$ is a CP
eigenstate the situation is far simpler. For  CP eigenstates
$CP\big|f_{CP}\big>=\pm\big|f_{CP}\big>$. It is useful to define
the amplitudes$ A=\big<f_{CP}\big|{\cal H}\big|B^0\big>,~~
\overline{A}=\big<f_{CP}\big|{\cal H}\big|\overline{B}^0\big>.$ If
$\left|{\overline{A}\over A}\right|\neq 1$, then we have ``direct"
CP violation in the decay amplitude, which we will discuss in
detail later. Here CP can be violated by having
\begin{equation}
\lambda = {q\over p}\cdot {\overline{A}\over A}\neq 1,
\end{equation}
which requires only that $\lambda$  acquire a non-zero phase, i.e.
$|\lambda|$ could be unity and CP violation can occur.

A comment on neutral $B$ production at $e^+e^-$ colliders is in
order. At the $\Upsilon (4S)$ resonance there is coherent
production of $B^0\bar{B}^0$ pairs. This puts the $B$'s in a
$C=-1$ state. In hadron colliders, or at $e^+e^-$ machines
operating at the $Z^0$, the $B$'s are produced incoherently. The
asymmetry is defined as
\begin{equation}
a_{f_{CP}}={{\Gamma\left(B^0(t)\to f_{CP}\right)-
\Gamma\left(\overline{B}^0(t)\to f_{CP}\right)}\over
{\Gamma\left(B^0(t)\to f_{CP}\right)+
\Gamma\left(\overline{B}^0(t)\to f_{CP}\right)}},
\end{equation}
which for $|q/p|=1$ gives
\begin{equation}
a_{f_{CP}}={{\left(1-|\lambda|^2\right)\cos\left(\Delta
mt\right)-2{\rm Im}\lambda \sin(\Delta mt)}\over {1+|\lambda|^2}}.
\label{eq:afcp}
\end{equation}
For the cases where there is only one decay amplitude $A$,
$|\lambda |$ equals 1, and we have
\begin{equation}
a_{f_{CP}}=-{\rm Im}\lambda \sin(\Delta mt).
\end{equation}
Only the amplitude, ${\rm -Im}\lambda$ contains information about
the level of CP violation, the sine term is determined only by $B_d$
mixing; the time integrated asymmetry is given by
\begin{equation}
a_{f_{CP}}=-{x \over {1+x^2}}{\rm Im}\lambda = -0.48 {\rm
Im}\lambda ~~. \label{eq:aint}
\end{equation}
This is quite lucky as the maximum size of the coefficient for any
$x$ is $-0.5$.

Let us now find out how ${\rm Im}\lambda$ relates to the CKM
parameters. Recall $\lambda={q\over p}\cdot {\overline{A}\over
A}$. The first term is the part that comes from mixing:
\begin{equation}
{q\over p}={{\left(V_{tb}^*V_{td}\right)^2}\over
{\left|V_{tb}V_{td}\right|^2}}
={{\left(1-\rho-i\eta\right)^2}\over
{\left(1-\rho+i\eta\right)\left(1-\rho- i\eta\right)}}
=e^{-2i\beta}{\rm~~and}
\end{equation}
\begin{equation}
{\rm Im}{q\over p}= -{{2(1-\rho)\eta}\over
{\left(1-\rho\right)^2+\eta^2}}=\sin(2\beta).
\end{equation}

To evaluate the decay part we need to consider specific final
states. Let's consider the final state $J/\psi K_s$. The decay
process proceeds via the diagram in Fig.~\ref{Bdiagrams2}(b),
where the $c\overline{c}$ forms a J/$\psi$. Here we do not get a
phase from the decay part because
\begin{equation}
{\overline{A}\over A} = {{\left(V_{cb}V_{cs}^*\right)^2}\over
{\left|V_{cb}V_{cs}\right|^2}}
\end{equation}
is real to order $1/\lambda^4$.

In this case the final state is a state of negative $CP$, i.e.
$CP\big|J/\psi K_s\big>=-\big|J/\psi K_s\big>$. This introduces an
additional minus sign in the result for ${\rm Im}\lambda$. Before
finishing discussion of this final state we need to consider in
more detail the presence of the $K_s$ in the final state. Since
neutral kaons can mix, we pick up another mixing phase (similar
diagrams as for $B^0$, see Fig.~\ref{Bdiagrams2}(e). This term
creates a phase given by
\begin{equation}
\left({q\over p}\right)_K={{\left(V_{cd}^*V_{cs}\right)^2}\over
{\left|V_{cd}V_{cs}\right|^2}},
\end{equation}
which is real to order $\lambda^4$. It is necessary to include this
term, however, since there are other  formulations of the CKM matrix
than Wolfenstein, which have the phase in a  different location. It
is important that the physics predictions not depend on the  CKM
convention.\footnote{Here we don't include CP violation in the
neutral kaon  since it is much smaller than what is expected in the
$B$ decay. The term of order $\lambda^4$ in $V_{cs}$ is necessary to
explain $K^0$ CP violation.}

\subsection{CP Violation Measurements}

\begin{figure}[htb]
\vspace{-15cm}
\includegraphics[width=0.6\textwidth,clip]{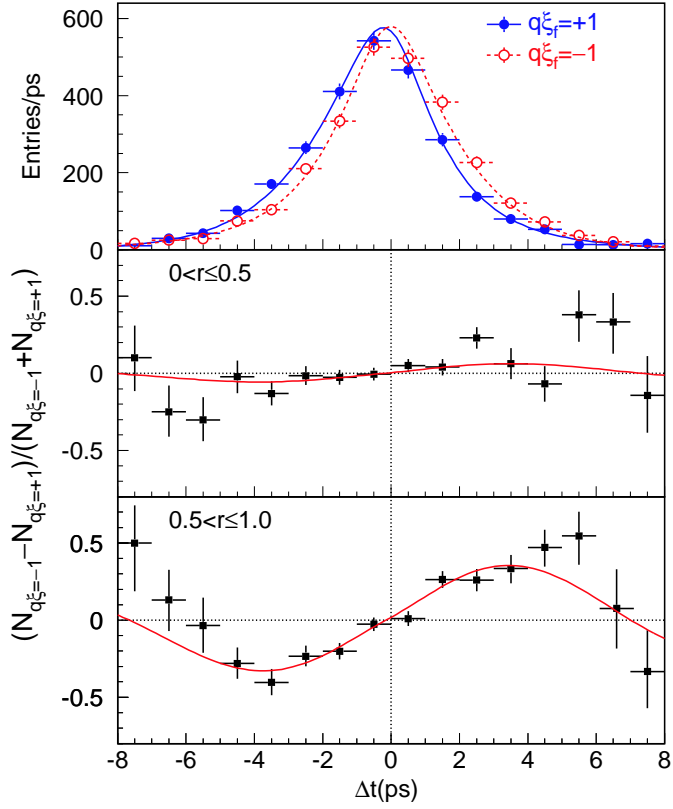}
 \caption{$\Delta t$ distributions from Belle for the events
    with $q\xi_f = -1$ (open points) and
    $q\xi_f = +1$ (solid points) with all modes combined (top),
    asymmetry between $q\xi_f=-1$ and $q\xi_f=+1$
    samples with $0 < r \le 0.5$ (middle), and
    with $0.5 < r \le 1$ (bottom). The variable $r$ refers to the
    probability of the correctness of the flavor tag, where $r$=1
    is almost assuredly correct, while $r$=0 conveys no
    information.
    The results of the global unbinned maximum-likelihood fit
    ($\sin(2\beta)$=0.728) are also shown.}
\label{fig:rawasym}
\end{figure}

The CP asymmetry $\sin(2\beta)$ has been measured by both Belle
and BaBar using both CP+ and CP- final states. Most of the latter
are $J/\psi K_S$, while most of the former are $J/\psi K_L$.
Fig.~\ref{fig:rawasym} shows the raw asymmetries and the fit
results for $(c\overline{c})K_S$ (top) and $J/\psi K_L$ (bottom)
\cite{Belle-psiKs}. The world average value of $\sin(2\beta)$ is
0.685$\pm$0.032 \cite{HFAG}.

The Belle collaboration pioneered the measurement of $\gamma$
using the charged decays $B^{\mp}\to D^0 K^{\mp}$, where the
$D^0\to K_S\pi^+\pi^-$. Here $D^0$ decays cannot be distinguished
from $\overline{D^0}$ decays, and they interfere. Measurements
from BaBar and Belle have been reported. BaBar averages in
additional information from $D^{*0}K^{\mp}$ and $D^{0}K^{0\mp}$,
finding $\gamma=(67 \pm 28 \pm 13\pm 11)^{\circ}$ and Belle,
omitting the last mode, obtains $\gamma=(67^{+14}_{-15} \pm 13\pm
11)^{\circ}$, where the last error is due the parametrization of
the $D^0$ decay Dalitz plot, and could be helped greatly by CLEO-c
measurements of the CP+ and CP- Dalitz plots \cite{Schune,Asner}.

\begin{figure}
\centerline{\epsfig{figure= 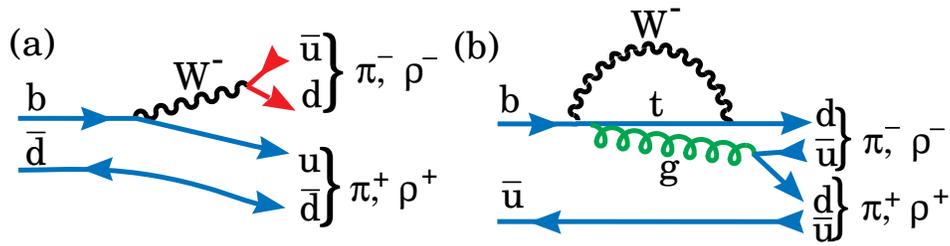,height=1.4in}}
\caption{Tree (a) and Penguin (b) processes for neutral $B$ decay
into either $\pi^+\pi^-$ or $\rho^+\rho^-$.}
\label{rhorho}       
\end{figure}

The angle $\alpha$ can be probed by measuring processes such as
$B^0\to\pi^+\pi^-$ or $\rho^+\rho^-$ as shown in
Fig.~\ref{rhorho}(a), because the combination of weak phases in the
mixing amplitude and the $b\to u$ decay amplitude are $\sin(2(\beta
+ \gamma)) = \sin(2(180-\alpha) =-\sin(2\alpha)$. Unfortunately, the
Penguin diagram in Fig.~\ref{rhorho}(b) has no weak phase and can be
significant in these processes. Thus the Penguin process can
``pollute" the measurement of $\alpha$ in these modes, but it can be
limited by using the upper limit on the branching ratio for
$B^0\to\rho^0\rho^0$ as shown by Grossman and Quinn \cite{GQ}. BaBar
first used this final state for CP violation measurement by showing
that it is almost fully polarized and that the Penguin term could be
usefully limited. Currently we have $\alpha=(96\pm 13 \pm
11)^{\circ}$, where the last error is due to the possible Penguin
contribution \cite{Schune}.

\section{Limits on New Physics}

\begin{figure}[htb]
\includegraphics[width=.6\textwidth]{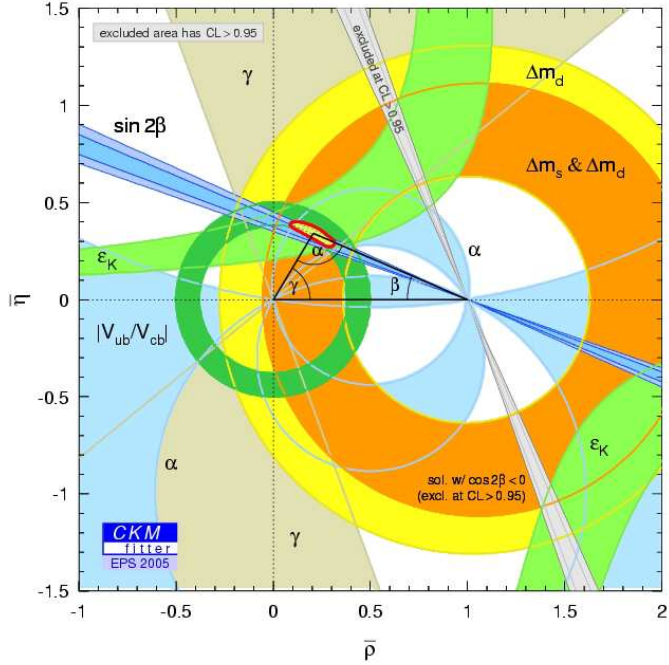} \\
\caption{Constraints on the $\rho$ and $\eta$ Wolfenstein parameters
after summer 2005, $\overline{\rho}=\rho(1-\lambda^2/2)$ and
$\overline{\eta}=\eta(1-\lambda^2/2)$.} \label{rhoeta}
\end{figure}

Constraints on the Wolfenstein $\rho$ and $\eta$ parameters are
given by many measurements and summarized in Fig.~\ref{rhoeta}
\cite{CKMfitter}. (For alternative fits see Ref.~\cite{utfits}.)
This plot is based on measurements of $|V_{ub}/V_{cb}|$,
$B^0-\overline{B}^0$ mixing, upper limits on $B_S$ mixing and the
CP violation measurements discussed here of $\alpha$, $\beta$ and
$\gamma$ as well as CP violation in the $K_L^0$ system.

\begin{figure}[htb]
\includegraphics[width=.5\textwidth]{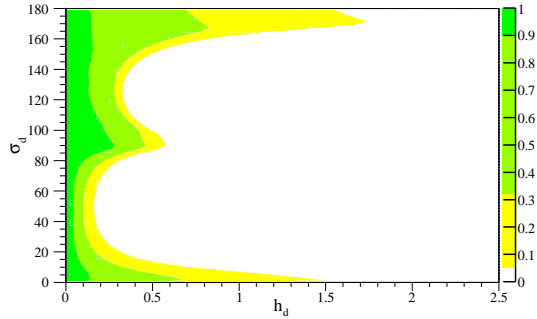}

\caption{The allowed range for $h_d$ and $\sigma_d$ from
\cite{NMFV}. The shaded scale on the right side indicates the
confidence level.} \label{fighsigma}
\end{figure}

Agashe \etal ~have established limits on New Physics (NP) arising
via $B^0$ mixing \cite{NMFV}, using a method that was modified from
that first used by Grossman \etal and Ligeti \cite{Ligeti}. They
assume that NP in tree level processes, such as those used to
measure $|V_{ub}|$ is negligible. They then parameterize NP in terms
of an amplitude $h$ and a phase $\sigma$ as
\begin{eqnarray}
 \Delta m_d&=&\left|1+h_d
e^{2i\sigma_d}\right| \Delta m_d^{\rm
  SM}\,,\qquad S_{\psi K}=\sin\left[2\beta+ 2\theta_d\right]~~,  \label{par}
\end{eqnarray}
where $\theta_d={\rm arg}\left(1+h_d
    e^{2i\sigma_d}\right)$. The CP asymmetry in $B^{\mp} \rightarrow DK^{\mp}$,
$A_{DK}$, is also a SM tree level transition and therefore is
unaffected by NP; $A_{DK}\sim \tan\gamma = \frac{ \eta }{ \rho }
$. Note that $A_{DK}$ depends only on $\rho$, $\eta$ in a
combination different than $V_{ ub }$. The CP asymmetry in $B^0\to
\rho^+\rho^-$, ${S_{\rho\rho}}$, is given by
${S_{\rho\rho}}\propto\sin\left( 2 \gamma + 2 \beta + 2 \theta_d
\right).$ Thus, $S_{ \rho \rho }$ also depends only on $\rho$,
$\eta$ {\em
  after} subtracting the phase of $B_d$ mixing (including the NP
phase) using $S^{ exp }_{ \psi K_s }$. Thus $\rho$, $\eta$ can be
determined even in presence of NP. The allowed size of NP admits a
range for $h_d$ of $h_d=0-0.4$,  for  $2\sigma_d=\pi-2\pi$. This
is demonstrated in Fig.~\ref{fighsigma} where the $h_d-\sigma_d$
allowed regions are shown.
Thus, the data do not yet exclude substantial contributions to NP
via $B_d$ mixing. In the case of NP via $B_S$ mixing, there are
almost no restrictions.

\begin{figure}[htb]
\includegraphics[width=.5\textwidth]{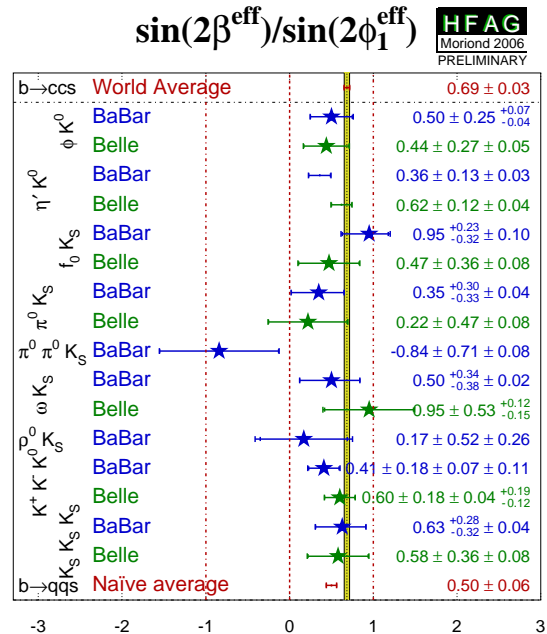}
\caption{Measurement of $\sin(2\beta)$ in Penguin dominated modes
versus that in $(c\overline{c})s$ modes. Note that $\sin(2\beta)$ is
sometimes called $\sin(2\phi_1)$. The superscript ``eff" indicates
that no attempt has been made to correct for the possible presence
of a $\cos(\Delta mt)$ term, see equation \ref{eq:afcp}.  }
\label{Penguin-CP}
\end{figure}

A hint of NP may be showing up in measurements of CP violation in
Penguin decays. A data summary is shown in Fig.~\ref{Penguin-CP}.
The trend is for these modes to have asymmetries below that in
$J/\psi K_S$ related modes. These modes may have additional
amplitudes, but calculations tend to show that these would result
in positive asymmetries, opposite to the observed effect
\cite{Benecke}. Each mode must be considered individually so
averaging them is not a reasonable approach.

\section{Future $B$ Decay Experiments}

The future of heavy physics may well be the provenance of
experiments at CERN starting in $\sim$2008 when significant data
will be taken by experiments at the LHC, a proton-proton collider
with 14 TeV of energy in the center-of-mass.

Three experiments are equipped to study $B$ decays. The LHCb
experiment is the only one specifically designed for this purpose.
The ATLAS and CMS experiments can, however, make some useful
measurements; they are intended to run a very high luminosity,
$10^{34}$cm$^{-2}$/s, while LHCb will run around $2\times
10^{32}$cm$^{-2}$/s. While CMS and ATLAS are designed to measure
new high mass particles in the central region, LHCb will detect
$b$-flavored hadrons produced in the forward direction along one
of the beams. The production mechanism tends to put both particles
in the detector acceptance, crucial for flavor tagging, i. e.
distinguish the flavor of the $b$'s at birth.

\begin{figure}[hbt]
\centerline{\epsfig{figure=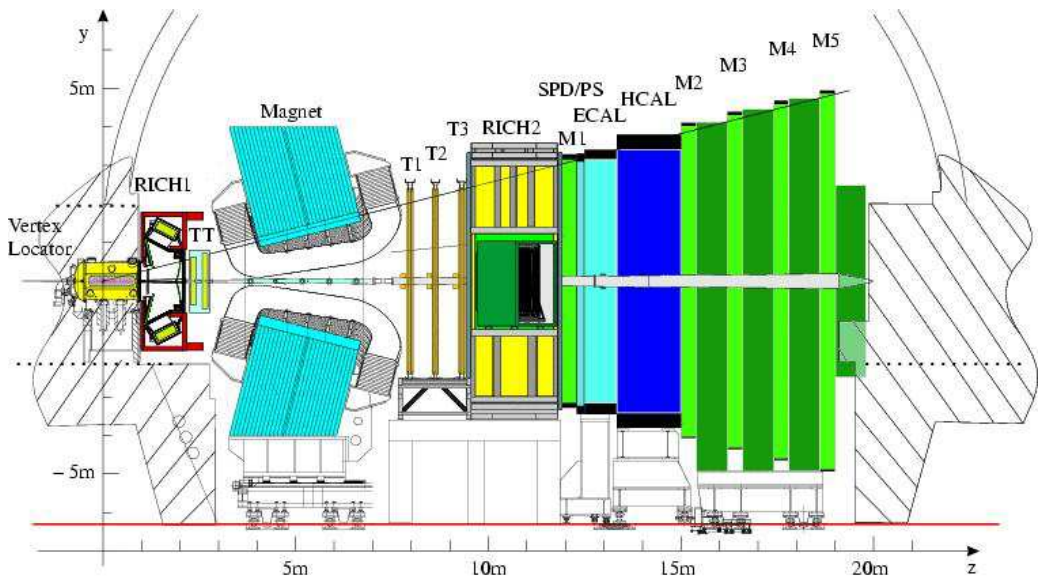,height=3in}}
\caption{\label{lhcb_det}A sketch of the LHCb detector showing the
Vertex Locator (VELO), the two RICH subsystems, the tracking
trigger stations (TT) before the magnet, the tracking stations
after the magnet (T1-T3), the Scintillating Pad detector (SPD),
the Prewshower (PS), the Electromagnetic Calorimeter (ECAL), the
Hadronic Calorimeter (HCAL) and the Muon Stations (M1-M5).}
\end{figure}

\begin{figure}[hbt]
\begin{tabular}{ccc}
\parbox{0.37\textwidth}{
\includegraphics[width=0.35\textwidth]{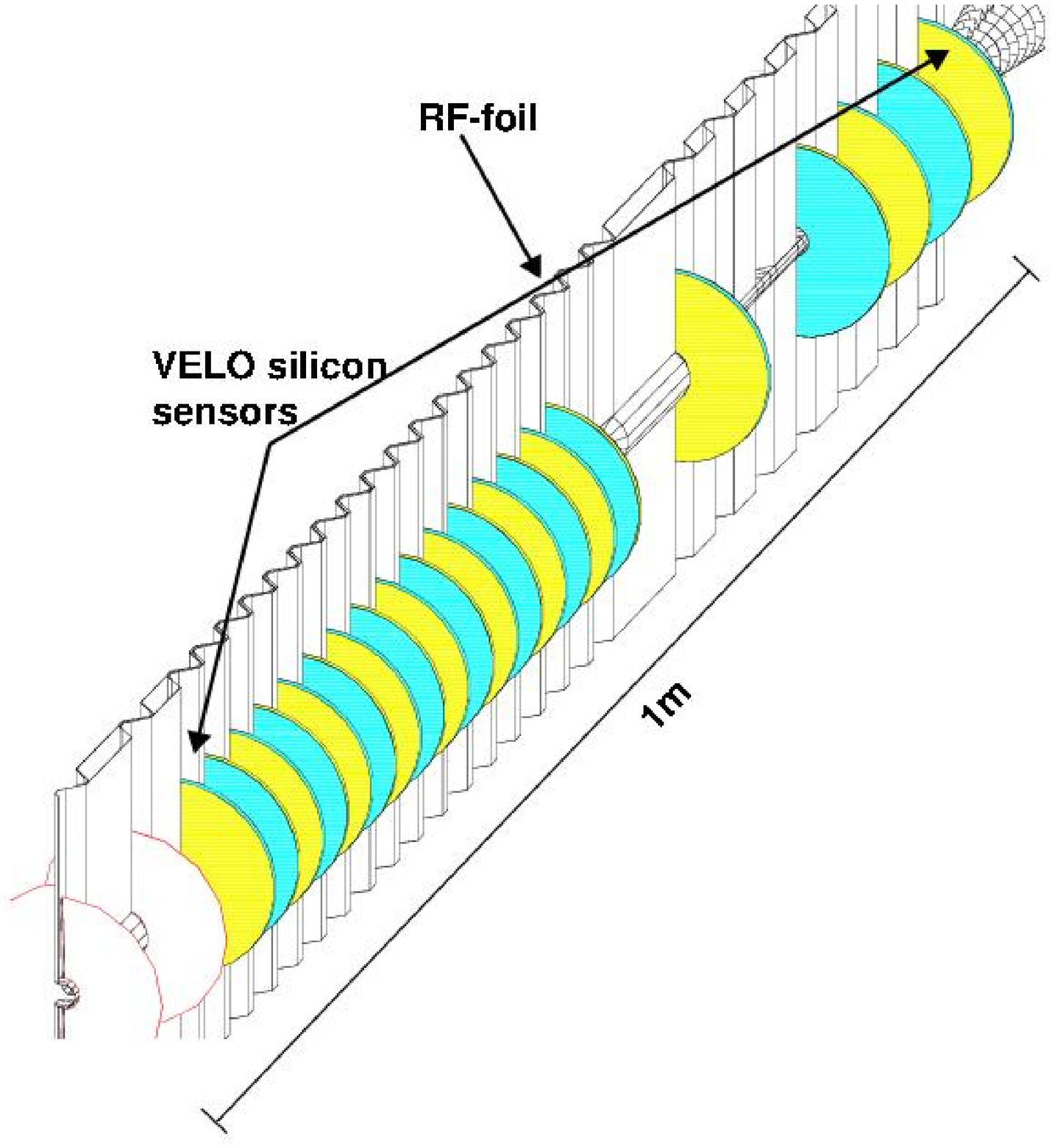}}
&
\parbox{0.37\textwidth}{
\includegraphics[width=0.35\textwidth]{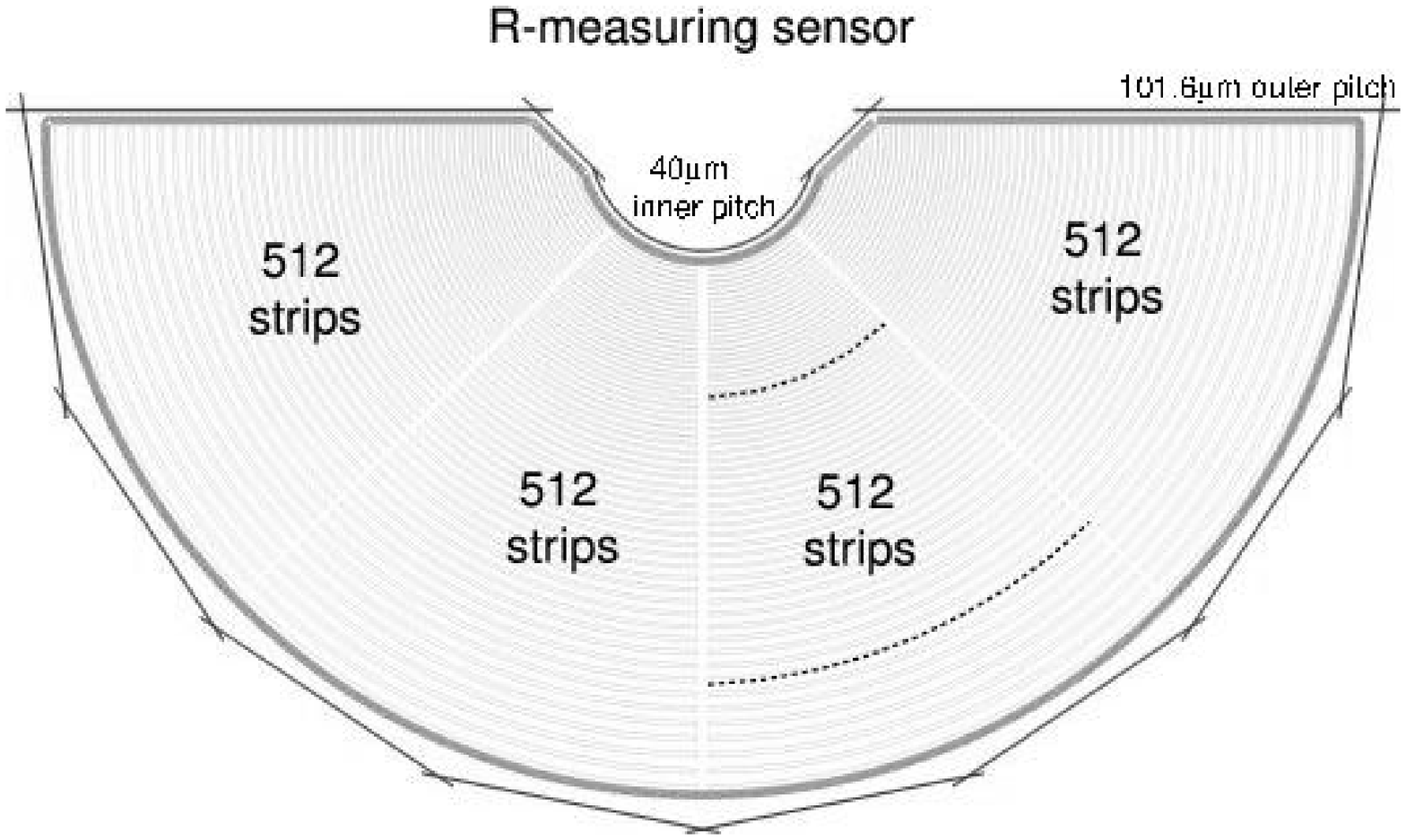}
\includegraphics[width=0.35\textwidth]{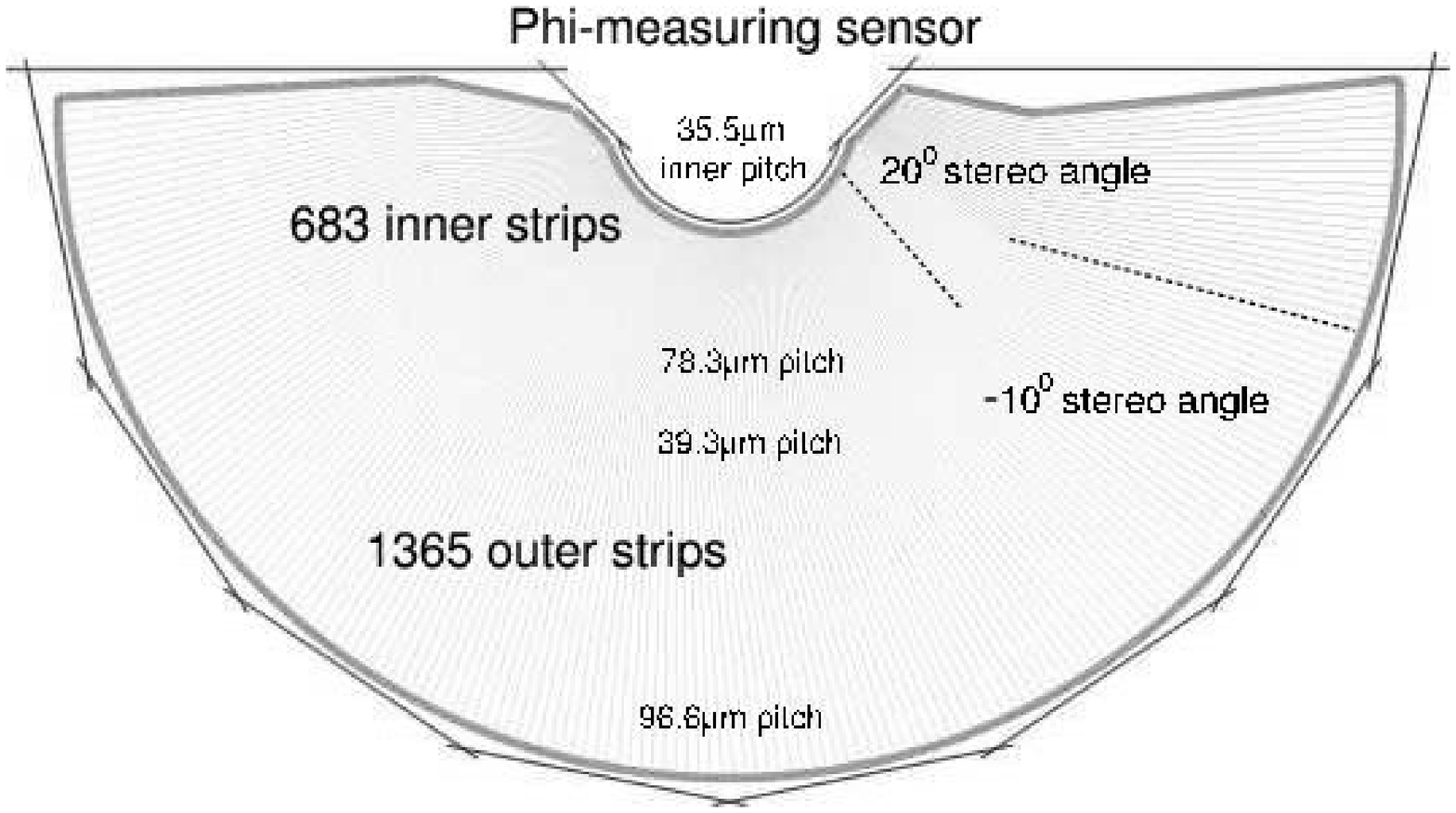}
} &
\parbox{0.22\textwidth}{
\rotatebox{90}{\includegraphics[height=0.22\textwidth]{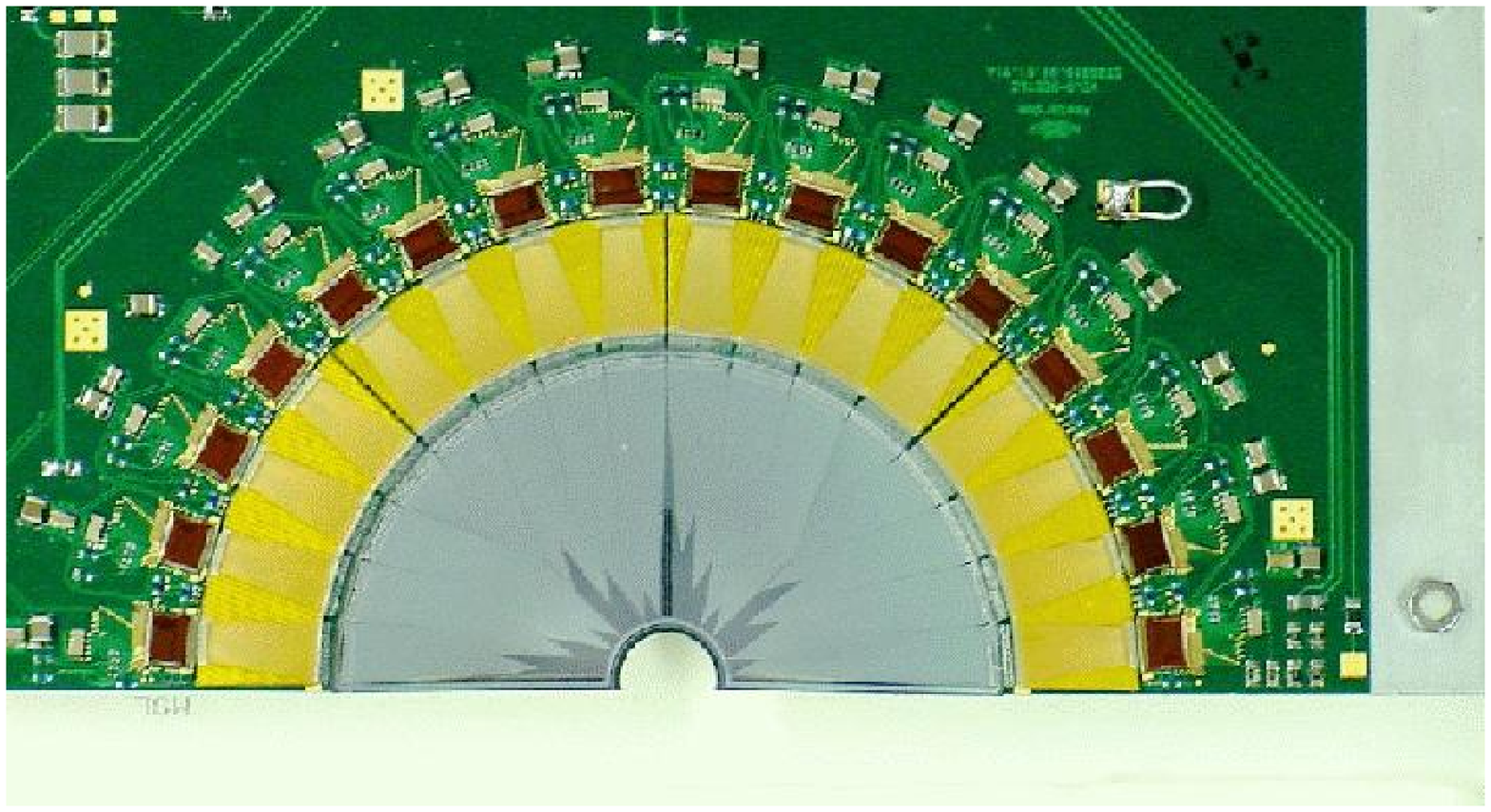}} }
\\
\parbox[t]{0.31\textwidth}{
(a) VELO with RF-foil,  21 $r-\phi$ detector stations, and two
upstream $r$ stations for the Pile Up system. } &
\parbox[t]{0.31\textwidth}{(b) $r$ and $\phi$ sensors. For each sensor, 2 readout strips are
indicated by dotted lines, for illustration.} &
\parbox[t]{0.18\textwidth}{
(c) Prototype Si sensor with readout electronics}
\end{tabular}
\caption{The LHCb Vertex Locator (VELO) \label{fig:Velo}}
\end{figure}

A sketch of the LHCb detector is shown in Fig.~\ref{lhcb_det}. A
silicon strip detector called ``VELO" is used to measure decay
vertices. The detectors are segmented along the radial and
azimuthal directions. The layout is shown Fig.~\ref{fig:Velo}, the
sensor geometry in (b) and a photograph in (c). There are two ring
imaging Cherenkov counters used to distinguish pions from kaons,
required because of the large range of momenta (1-100 GeV/c) that
occur. An electromagnetic calorimeter constructed from
scintillating fibers and lead detects $\gamma$'s $\pi^0$'s and
$\eta$'s; it also identifies electrons. The iron filter after the
hadron calorimeter (HCAL) interspersed with the chambers M2-M5 is
used to identify muons.  The calorimetry, both electromagnetic and
hadronic provide real time information used in the first trigger
level (called Level 0) for charged particles or neutral energy at
transverse momenta that are likely to come from $b$ decays. A
``pile-up" device is also used to identify beam crossings with
more than one interaction. More details about the detector can be
found in \cite{LHCb}.

%


The KEK accelerator has produced very impressive luminosities and
there are plans to improve it. This concept is called
``Super-Belle." There would be both machine and detector
improvements allowing running up to an instantaneous luminosity of
$\sim 5\times 10^{35}$cm$^{-2}$/s. Currently, this is a proposal
that has yet to be acted on. Another similar proposal was also
formulated by the ``Super-BaBar" group. It however has not been
supported by SLAC or the U. S. Dept. of Energy.

A group in Frascati, Italy has been exploring the possibility of
using recirculating electron linacs as the basis of a novel
$e^+e^-$ collider in the Upsilon region \cite{Frascati}. This
machine would not have appreciable synchrotron radiation, so
current detector technologies would work just fine. However, the
number of interactions per crossing could be large.




\section{Conclusions}

The study of the decays of $b$-flavored hadrons has advanced
greatly from its early beginnings. We have one precision
measurement, namely that of $\sin(2\beta)$ and initial
measurements of $\alpha$ and $\gamma$. Yet much more needs to be
done. $|V_{ub}|$ needs to be made more precise by improvements in
QCD calculations and comparisons with charmless semileptonic
decays in the appropriate kinematic regions suitable for reliable
theoretical predictions. Measurements of CP violation in $B_S$
decays are of prime importance. After the termination of current
experimental efforts in flavor physics in the U. S. at the end of
this decade, experimental progress will depend on experiments at
the LHC, in particular LHCb, and at Belle or a possible
Super-Belle in Japan. These experiments will be essential in
interpreting the New Physics we expect to find at the LHC.

\begin{theacknowledgments}
This work was supported by the U. S. National Science Foundation
under grant \#0353860. I thank A. Alejandro and A. Bashir for making
the  X Mexican Workshop on Particles and Fields a pleasant and
useful meeting. I also had useful discussions with K. Agashe and M.
Artuso.
\end{theacknowledgments}




\end{document}